\documentclass[11pt]{article}
\usepackage{array,ifthen,amsmath,amssymb,amsthm,latexsym,epsfig,bm,verbatim}

\newcommand{\CR}{\noindent\hrule}
\newcommand{\MCR}{\vskip 0.2cm\CR\vskip 0.2cm}

\def\eps{\varepsilon}

\def\R{{\mathbb R}}

\def\<{\langle}
\def\>{\rangle}
\def\({\left(} 
\def\){\right)} 

\newtheorem{thm}{Theorem}[section]
\newtheorem{pro}[thm]{Proposition}
\newtheorem{obs}[thm]{Observation}
\newtheorem{exc}{Exercise}
\newtheorem{nte}{Note}
\newtheorem{rem}[thm]{Remark}
\newtheorem{lem}[thm]{Lemma}
\newtheorem{cor}[thm]{Corollary}

\newtheorem{con}{Conjecture}
\newtheorem{exm}[thm]{Example}
\newtheorem{dfn}[thm]{Definition}
\newtheorem{fct}[thm]{Fact}
\newtheorem{que}[thm]{Question}

\newenvironment{prf}[1]{\noindent{\bf{Proof #1\\}}}{$\hfill\blacksquare$\nopagebreak[4]\vskip 0.3cm}

\def\BS{\begin{footnotesize}}
\def\ES{\end{footnotesize}}
\def\BCON{\begin{con}}
\def\ECON{\end{con}}
\def\BOBS{\begin{obs}}
\def\EOBS{\end{obs}}

\def\BPRO{\begin{pro}}
\def\EPRO{\end{pro}}
\def\BIDE{\begin{ide}}
\def\EIDE{\end{ide}}
\def\BTHM{\begin{thm}}
\def\ETHM{\end{thm}}
\def\BDEF{\begin{dfn}\rm}
\def\EDEF{\end{dfn}}
\newcommand\BPRF[1][:]{\begin{prf}{#1}}
\def\EPRF{\end{prf}}
\def\BLEM{\begin{lem}}
\def\ELEM{\end{lem}}
\def\BEX{\begin{exm}\rm}
\def\EEX{\end{exm}}
\def\BEXC{\begin{exc}\rm}
\def\EEXC{\end{exc}}
\def\BCOR{\begin{cor}}
\def\ECOR{\end{cor}}
\def\BQUE{\begin{que}}
\def\EQUE{\end{que}}
\newcommand\BSOL[1][:]{\begin{sol}{#1}}
\def\ESOL{\end{sol}}
\def\BNTE{\begin{nte}}
\def\ENTE{\end{nte}}

\def\BIT{\begin{itemize}}
\def\EIT{\end{itemize}}
\def\BREM{\begin{rem}\rm}
\def\EREM{\end{rem}}
\def\BC{\begin{center}}
\def\EC{\end{center}}
\def\BEQ{\begin{equation}}
\def\EEQ{\end{equation}}

\date{}

\title{Approximating $L_1$-distances between mixture distributions using random projections}

\author{Satyaki Mahalanabis $\quad\quad\quad\quad\quad\quad$ Daniel \v{S}tefankovi\v{c}
\vspace{0.2cm}\\
Department of Computer Science\\ University of Rochester\\ Rochester, NY 14627\\
{\tt \{smahalan,stefanko\}@cs.rochester.edu}}

\begin{document}

\maketitle

\begin{abstract}
We consider the problem of computing $L_1$-distances between every pair of
probability densities from a given family. We point out that the technique
of Cauchy random projections~\cite{I06} in this context turns into stochastic
integrals with respect to Cauchy motion.

For piecewise-linear densities these integrals can be sampled from
if one can sample from the stochastic integral of the function
$x\mapsto (1,x)$. We give an explicit density function for this
stochastic integral and present an efficient (exact) sampling
algorithm. As a consequence we obtain an efficient algorithm to
approximate the $L_1$-distances with a small relative error.

For piecewise-polynomial densities we show how to approximately
sample from the distributions resulting from the stochastic
integrals. This also results in an efficient algorithm to
approximate the $L_1$-distances, although our inability to get
exact samples worsens the dependence on the parameters.
\end{abstract}

\section{Introduction}

Consider a finite class ${\cal F} = \{ f_1, f_2, \ldots, f_m \}$ of probability densities.
We want to compute the distance between every pair of members of ${\cal F}$.
We are interested in the case where each member of ${\cal F}$ is a mixture of finitely
many probability density functions, each having a particular functional form
(e.\,g., uniform, linear, exponential, normal, etc.). Such classes of distributions are frequently
encountered in machine learning (e.\,g., mixture models, see~\cite{Bish06}) and  nonparametric density
estimation (e.\,g., histograms, kernels, see~\cite{DG01}). The number of distributions in a mixture
gives a natural measure of complexity which we use to express the running time of our
algorithms.

For some classes of distributions exact algorithms are possible, for example, if each distribution in
${\cal F}$ is a piecewise linear function consisting of $n$ pieces then we can compute the distances
between all pairs in time $\Theta(m^2 n)$. For other classes of distributions (for example,
mixtures of normal distributions) exact computation of the distances might not be possible.
Thus we turn to randomized approximation algorithms. A {\em $(\delta,\eps)$-relative-error
approximation scheme} computes $D_{jk},\ j,k\in [m]$ such that with probability at least $1-\delta$ we have
\begin{equation*}\label{rela}
(\forall j,k\in [m])\ \ (1 - \eps) D_{jk} \leq \| f_j - f_k \|_1 \leq (1 + \eps) D_{jk}.
\end{equation*}
A {\em $(\delta,\eps)$-absolute-error approximation scheme} computes $D_{jk},\ j,k\in [m]$ such that
with probability at least $1-\delta$ we have
\begin{equation*}\label{abso}
(\forall j,k\in [m])\ \ D_{jk} - \eps \leq \| f_j - f_k \|_1 \leq D_{jk} + \eps.
\end{equation*}

A direct application of the Monte Carlo method (\cite{MU49}, see \cite{M87}) immediately yields the following
absolute-error approximation scheme. Let $X_{jk}$ be sampled according to $f_j$ and let
$Y_{jk}={\rm sgn}(f_j(X_{jk})-f_k(X_{jk}))$, where ${\rm sgn}:\R\rightarrow\{-1,0,1\}$ is
the sign function. The expected value of $Y_{jk}+Y_{kj}$ is equal to $\|f_j-f_k\|_1$, indeed
$$
E[Y_{jk}+Y_{kj}]=\int (f_j(x)-f_k(x))\,{\rm sgn}(f_j(x)-f_k(x))\,{\rm d}x=\|f_j-f_k\|_1.
$$
Thus, to obtain a $(\delta,\eps)$-absolute-error approximation scheme it is enough to
approximate each $Y_{jk}$ with absolute error $\eps/2$ and confidence $1-\delta/m^2$.
By the Chernoff bound $O(\eps^{-2}\ln(m^2/\delta))$ samples from each $Y_{jk}$ are enough.
(The total number of samples from the $f_j$ is $O(m \eps^{-2}\ln(m^2/\delta)$,
since we can use the same sample from $f_j$ for $Y_{j1},\dots,Y_{jm}$. The total
number of evaluations is $O(m^2 \eps^{-2}\ln(m^2/\delta)$.) The running time of this
algorithm will compare favorably with the exact algorithm if {\em sampling} from the densities
and {\em evaluation} of the densities at a point can be done fast.
(For example, for piecewise linear densities both sampling and evaluation can be done
in $O(\log n)$ time, using binary search.)
Note that the evaluation oracle is essential (cf. \cite{BFRSW00} who only allow use of sampling oracles).

In the rest of the paper we will focus on the harder relative-error approximation
schemes (since the $L_1$-distance between two distributions is at most $2$,
a relative-approximation scheme immediately yields an absolute-error approximation scheme).
Our motivation comes from an application (density estimation) which requires a
relative-error scheme~\cite{MS08}.

Now we outline the rest of the paper. In Section~\ref{z2} we review Cauchy random projections;
in Section~\ref{z3} we point out that for density functions Cauchy random projections become stochastic
integrals; in Section~\ref{z4} we show that for piecewise linear functions we can sample from these
integrals (using rejection sampling, with bivariate student distribution as the envelope) and
as a consequence we obtain efficient approximation algorithm for relative-error all-pairs-$L_1$-distances.
Finally, in Section~\ref{z5}, we show that for piecewise polynomial functions one can
approximately sample from the integrals, leading to slightly less efficient approximation algorithms.

\section{Cauchy random projections}\label{z2}

Dimension reduction (the most well-known example is the Johnson-Lindenstrauss lemma
for $L_2$-spaces~\cite{JL84}) is a natural technique to use here. We are interested in
$L_1$-spaces for which the
analogue of the Johnson-Lindenstrauss lemma is not possible~\cite{BC05,LN04} (that is,
one cannot project points into a low dimensional $L_1$-space and preserve distances
with a small relative error). However one can still project points to short vectors from
which $L_1$-distances between the original points can be approximately recovered
using {\em non-linear} estimators~\cite{LHC07,I06}.

A particularly fruitful view of the dimensionality ``reduction'' (with non-linear estimators)
is through stable distributions (\cite{JS82,I06}): given vectors $v_1,\dots,v_m$ one defines
(dependent) random variables $X_1,\dots,X_m$ such that the distance of $v_j$ and $v_k$ can be
recovered from $X_j-X_k$ (for all $j,k\in [m]$). For example, in the case of $L_1$-distances
$X_j-X_k$ will be from Cauchy distribution $C(0,\|v_j-v_k\|_1)$, and hence the recovery problem is
to estimate the scale parameter $R$ of Cauchy distribution $C(0,R)$. This
is a well-studied problem (see, e.\,g.,~\cite{HBA70}). We can, for example, use the following nonlinear
estimator (other estimators, e.\,g., the median are also possible~\cite{I06}):

\BLEM[Lemma~7 of ~\cite{LHC07}]\label{l:lhctail}
Let $X_1, X_2, \ldots, X_t$ be independent samples from the Cauchy distribution $C(0,D)$.
Define the geometric mean estimator without bias-correction $\hat{D}_{gm}$ as
\[ \hat{D}_{gm} = \prod_{j=1}^{t} |X_j|^{1/t}.\]
Then for each $\eps\in [0,1/2]$, we have
\begin{eqnarray*}
P\(\hat{D}_{gm} \in [ (1-\eps)D, (1+\eps)D ] \) \geq 1 - 2\exp(-t\eps^2/8).
\end{eqnarray*}
\ELEM

We first illustrate how Cauchy random projections immediately give an efficient relative-error approximation
scheme for piecewise uniform distributions.

Let ${\cal F}$ consist of $m$ piecewise uniform densities, that is, each member of ${\cal F}$ is
a mixture of $n$ distributions each uniform on an interval. Let $a_1,\dots,a_{s}$ be the
endpoints of all the intervals that occur in ${\cal F}$ sorted in the increasing order
(note that $s\leq 2mn$). Without loss of generality, we can assume that
each distribution $f_j\in {\cal F}$ is specified by $n$ pairs
$(b_{j1},c_{j1}),\dots,(b_{jn},c_{jn})$ where
$1\leq b_{j1}<c_{j1}<\dots<b_{jn}<c_{jn}\leq s$,
and for each pair $(b_{j\ell},c_{j\ell})$ we are also given
a number $\alpha_{j\ell}$ which is the value of $f_j$ on
the interval $[a_{b_{j\ell}},a_{c_{j\ell}})$.

Now we will use Cauchy random projections to compute the pairwise $L_1$-distances between the $f_j$
efficiently. For $\ell\in\{1,\dots,s-1\}$ let $Z_\ell$ be independent from the
Cauchy distribution $C(0,a_{\ell+1}-a_\ell)$. Let $Y_{\ell}=Z_1+\dots+Z_{\ell-1}$, for $\ell=1,\dots,s$.
Finally, let
\begin{equation}\label{stint}
X_j:=\sum_{\ell=1}^n \alpha_{j\ell} (Y_{c_{j\ell}} - Y_{b_{j\ell}})=\sum_{\ell=1}^n\alpha_{j\ell}
(Z_{b_{j\ell}}+\dots+Z_{c_{j\ell}-1}).
\end{equation}
Note that $X_j$ is a sum of Cauchy random variables and hence has Cauchy distribution (in fact it
is from $C(0,1)$). Thus $X_j-X_k$ will be from Cauchy distribution as well. The coefficient of
$Z_\ell$ in $X_j-X_k$ is the difference of $f_j$ and $f_k$ on interval $[a_\ell,a_{\ell+1})$.
Hence the contribution of $Z_{\ell}$ to $X_j-X_k$ is from Cauchy distribution
$C(0,\int_{a_{\ell}}^{a_{\ell+1}} |f_j(x)-f_k(x)|\,{\rm d}x)$, and thus
$X_j-X_k$ is from Cauchy distribution $C(0,\|f_j-f_k\|_1)$.

\BREM\label{det}
In the next section we will generalize the above approach to piecewise degree-$d$-polynomial
densities. In this case for each $(b_{j\ell},c_{j\ell})$ we are given a vector
$\alpha_{j\ell}\in\R^{d+1}$ such that the value of $f_j$ on interval $[a_{b_{j\ell}},a_{c_{j\ell}})$
is given by the following polynomial (written as an inner product):
$$
f_j(x)=(1,x,\dots,x^d)\cdot \alpha_{j\ell}.
$$
\EREM

\section{Cauchy motion}\label{z3}

A natural way of generalizing the algorithm from the previous section to arbitrary
density functions is to take infinitesimal intervals. This leads one to the well-studied area of stochastic
integrals w.r.t. symmetric $1$-stable L\'evy motion (also called Cauchy motion). Cauchy motion
is a stochastic process $\{X(t),t\in\R\}$ such that $X(0)=0$, $X$ has independent increments
(i.\,e., for any $t_1\leq t_2\leq\dots\leq t_k$ the random variables
$X(t_2)-X(t_1),\dots,X(t_k)-X(t_{k-1})$ are independent), and $X(t)-X(s)$ is
from Cauchy distribution $C(0,|t-s|)$. Intuitively, stochastic integral of a {\em deterministic function}
w.r.t. Cauchy motion is like a regular integral, except one uses $X(t)-X(s)$ instead of $t-s$
for the length of an interval (see section 3.4 of~\cite{ST94} for a readable formal
treatment).

We will only need the following basic facts about
stochastic integrals of deterministic functions w.r.t. Cauchy motion (which we will denote ${\rm d}{\cal L}(x)$),
see~\cite{ST94}, Chapter 3.

\begin{fct}\label{fact1}
Let $f:\R\rightarrow\R$ be a (Riemann) integrable function. Let
$X=\int_a^b f(x)\,{\rm d}{\cal L}(x)$.
Then $X$ is a random variable from Cauchy distribution $C(0,R)$ where
\begin{equation}\label{mrch}
R=\int_a^b |f(x)|\,{\rm d}x.
\end{equation}
\end{fct}

\begin{fct}\label{fact2}
Let $f_1,\dots,f_d:\R\rightarrow\R$ be (Riemann) integrable functions. Let $\phi=(f_1,\dots,f_d):\R\rightarrow\R^d$.
Let $(X_1,\dots,X_d)=\int_a^b \phi(x)\,{\rm d}{\cal L}(x)$.
Then $(X_1,\dots,X_d)$ is a random variable with characteristic function
$$
\hat{f}(c_1,\dots,c_d)=\exp\(-\int_{a}^b |c_1f_1(x)+\dots+c_df_d(x)|\,{\rm d}x\).
$$
\end{fct}

\begin{fct}\label{fact3}
Let $f,g:\R\rightarrow\R$ be (Riemann) integrable functions. Let $a < b,\alpha,\beta\in R$. Then
$$
\int_a^b (\alpha f+\beta g)\,{\rm d}{\cal L}(x) =
\alpha\int_a^b f\,{\rm d}{\cal L}(x) + \beta\int_a^b g\,{\rm d}{\cal L}(x).
$$
Let $h(x)=f(a+(b-a)x)$. Then
$$
\int_a^b f(x) \,{\rm d}{\cal L}(x) = (b-a)\int_0^1 h(x) \,{\rm d}{\cal L}(x).
$$
\end{fct}

From facts~\ref{fact1} and \ref{fact3} it follows that the problem of approximating the $L_1$-distances
between densities can be solved if we can evaluate stochastic integrals w.r.t. Cauchy motion;
we formalize this in the following observation.

\BOBS
Let $f_1,\dots,f_m:\R\rightarrow\R$ be probability densities. Let $\phi:\R\rightarrow\R^m$ be defined by
$\phi(x)=(f_1(x),\dots,f_m(x))$. Consider
\begin{equation}\label{pakoq}
(X_1,\dots,X_m)=\int_{-\infty}^{\infty} \phi(x)\,{\rm d}{\cal L}(x).
\end{equation}
For all $j,k\in [m]$ we have that $X_j-X_k$ is from Cauchy distribution $C(0,\|f_j-f_k\|_1)$.
\EOBS

Note that the $X_j$ defined by \eqref{stint} are in fact computing the integral in~\eqref{pakoq}. For
piecewise uniform densities it was enough to sample from the Cauchy distribution to compute the integral.
For piecewise degree-$d$-polynomial densities it will be enough to sample from the following distribution.

\BDEF
Let $\phi:\R\rightarrow\R^{d+1}$ be defined by $\phi(x)=(1,x,x^2,\dots,x^d)$. Let ${\rm CI}_d(a,b)$
be the distribution of $Z$, where
$$
Z:=(Z_0,\dots,Z_d):=\int_{a}^b \phi(x)\,{\rm d}{\cal L}(x).
$$
\EDEF

Note that given a sample from ${\rm CI}_d(0,1)$, using $O(d^2)$ arithmetic operations
we can obtain a sample from ${\rm CI}_d(a,b)$, using Fact~\ref{fact3}.

\BLEM\label{lpiece}
Let ${\cal F}$ consist of $m$ piecewise degree-$d$-polynomial densities, each consisting of $n$
pieces (given as in Remark~\ref{det}). Let $t\geq (8/\eps)^2\ln (m^2/\delta)$ be an integer.
Assume that we can sample from ${\rm CI}_d(0,1)$ using $T_d$ operations. We can obtain
$(\delta,\eps)$-relative-error approximation of $L_1$-distances between all
pairs in~${\cal F}$, using $O((d^2+T_d) mnt+m^2t)$ arithmetic operations.
\ELEM

\BPRF
For $\ell\in\{1,\dots,s-1\}$ let $Z_\ell$ be independent from ${\rm CI}_d(a_\ell,a_{\ell+1})$ distribution. Let
$Y_{\ell}=Z_1+\dots+Z_{\ell-1}$, for $\ell=1,\dots,s$. Finally, for each $j\in [m]$, let
\begin{equation}\label{stint2}
X_j:=\sum_{\ell=1}^n \alpha_{j\ell}\cdot (Y_{c_{j\ell}} - Y_{b_{j\ell}})=\sum_{\ell=1}^n\alpha_{j\ell}
\cdot (Z_{b_{j\ell}}+\dots+Z_{c_{j\ell}-1}).
\end{equation}
Note that $Y_{c_{j\ell}} - Y_{b_{j\ell}}$ is from $C(a_{b_{j\ell}},a_{c_{j\ell}})$
and hence
$$
\alpha_{j\ell}\cdot (Y_{c_{j\ell}} - Y_{b_{j\ell}}) = \int_{a_{b_{j\ell}}}^{a_{c_{j\ell}}} f_j(x)\,{\rm d}{\cal L}(x).
$$
Thus $(X_1,\dots,X_m)$ defined by \eqref{stint2} compute \eqref{pakoq}.

For every $j,k\in [m]$ we have that $X_j-X_k$ is from Cauchy distribution
\begin{equation*}\label{werty}
C(0,\|f_j-f_k\|_1).
\end{equation*}
If we have $t$ samples from each $X_1,\dots,X_m$ then using Lemma~\ref{l:lhctail} and union bound with
probability $\geq 1-\delta$ we recover all $\|f_j-f_k\|_1$ with relative error $\eps$.

Note that $s\leq 2mn$ and hence for the $Z_\ell$ we used $\leq 2mnt$ samples from ${\rm CI}(0,1)$
distribution, costing us $O((d^2+T_d) mnt)$ arithmetic operation. Computing the $Y_\ell$ takes $O(mnt)$ operations.
Computing the $X_j$ takes $O(mnt)$ operations. The final estimation of the distances
takes $O(m^2 t)$ operations.
\EPRF

\section{Piecewise linear functions}\label{z4}

The density function of ${\rm CI}_1(0,1)$ can be computed explicitly, using the inverse Fourier transform;
the proof is deferred to the appendix. The expression for the density allows us to construct
efficient sampling algorithm, which in turn yields an efficient approximation algorithm for
all-pairs-$L_1$-distances for piecewise linear densities. We obtain the following result.

\BTHM\label{tele}
Let ${\cal F}$ consist of $m$ piecewise linear densities, each consisting of $n$
pieces (given as in Remark~\ref{det}). We can obtain
$(\delta,\eps)$-relative-error approximation of $L_1$-distances between all
pairs in~${\cal F}$, using $O(m(m+n)\eps^{-2}\ln(m/\delta))$ arithmetic operations.
\ETHM

Now we state the density of ${\rm CI}_1(0,1)$. In the following $\Re(x)$ denotes the
real part of a complex number $x$.

\BTHM\label{looo}
Let $\phi:{\mathbb R}\rightarrow{\mathbb R}^2$ be the function $\phi(x)=(1,x)$. Let
$$Z=(X_1,X_2)=\int_0^1\phi(z)\,{\rm d}{\cal L}(z).$$
For $x_1\neq 2x_2$ the density function of $Z$ is given by
\begin{equation}\label{eden}
f(x_1,x_2)=
\frac{4/\pi^2}{1 + 6 x_1^2 + x_1^4 - 16 x_1 x_2 + 16 x_2^2}
+\frac{2}{\pi^2}\,\Re\(\frac{{\rm atan}(iQ/(x_1-2x_2))}
{Q^{3/2}}\),
\end{equation}
where
\begin{equation}\label{qeq}
Q=1 - 2 i x_1  + x_1^2 + 4 i x_2.
\end{equation}
For $x_1=2x_2$ the density is given by
\begin{equation}\label{eden2}
f(x_1,x_2)=
\frac{4/\pi^2}{(1 + x_1^2)^2}
+\frac{1}{\pi (1+x_1^2)^{3/2}}.
\end{equation}
\ETHM

\begin{figure}[htb]
    \begin{center}
        \includegraphics[type=eps,ext=.eps,read=.eps,angle=0,scale=1]{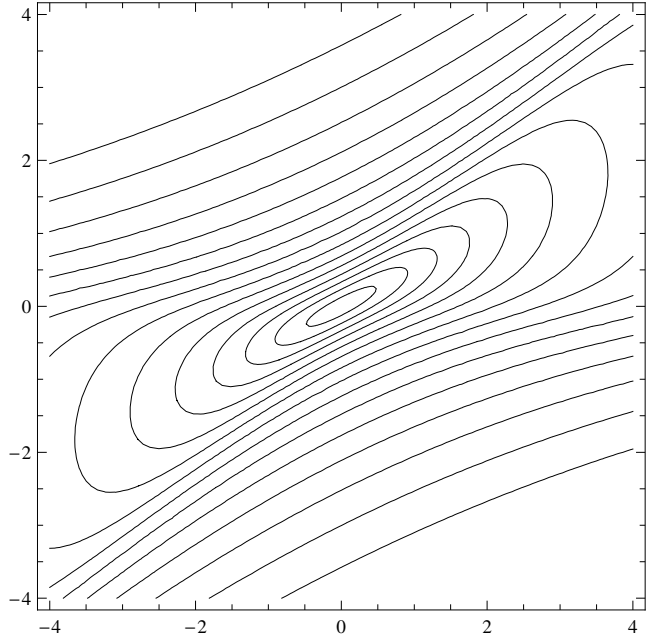}
      \caption{The density plot of $(X_1,X_2)=\int_{0}^1 (1,z)\,{\rm d}{\cal L}(z)$.
      The contours are at levels $2^{-15},2^{-14},\dots,2^{-1}$.
      }
      \label{fig:piecewise}
    \end{center}
\end{figure}

Next we show how to efficiently sample from the ${\rm CI}_1(0,1)$ distribution
by rejection sampling using the bivariate student distribution
as the envelope.

Let $\Sigma$ be a positive-definite $2\times 2$ matrix. The bivariate
student distribution with $1$ degree of freedom is given by the following formula
(see, e.\,g.,~\cite{ES00}, p.~50)
$$
g({\mathbf x})=\frac{|{\rm det}(\Sigma)|^{-1/2}}{2\pi}\(1+\frac{{\mathbf x}^T \Sigma^{-1} {\mathbf x}}{2}\)^{-3/2}.
$$
It is well-known how to sample $X$ from this distribution: let $X=\Sigma^{1/2} Y /\sqrt{W}$,
where $Y,W$ are independent with $Y\sim N_2(0,I)$ (the two dimensional gaussian) and
$W\sim\chi^2(1)$ (chi-squared distribution with $1$ degree of freedom).

We are going to use the bivariate student distribution with the following density
\begin{equation}\label{ZZZ}
g({\mathbf x})=\frac{1}{\pi}\(1+x_1^2+(2x_2-x_1)^2\)^{-3/2}.
\end{equation}

We show that the density function of the ${\rm CI}_1(0,1)$ distribution
is bounded by a constant multiple of \eqref{ZZZ} (the proof is deferred to
the appendix).

\BLEM\label{kkkk}
Let $f({\mathbf x})$ be given by \eqref{eden} and \eqref{eden2}. Let
$g({\mathbf x})$ be given by \eqref{ZZZ}. For every ${\mathbf x}\in\R^2$ we
have
\begin{equation*}\label{BOO}
f({\mathbf x})\leq \frac{C}{\pi}\cdot g({\mathbf x}),
\end{equation*}
where $C=25$.
\ELEM

As an immediate corollary of Lemma~\ref{kkkk} we obtain an efficient sampling
algorithm for ${\rm CI}_1(0,1)$ distribution, using rejection sampling (see, e.\,g.,
~\cite{ES00}).

\BCOR\label{cansample}
There is a sampler from ${\rm CI}_1(0,1)$ which uses
a constant number of samples from from $N(0,1)$ and $\chi^2(1)$
(in expectation).
\ECOR

\BPRF[of Theorem~\ref{tele}:]
The theorem follows from Corollary~\ref{cansample} and Lemma~\ref{lpiece}.
\EPRF

\BREM
Lemma~\ref{kkkk} is true with $C=\pi 2^{3/2}$ (we skip the technical proof).
The constant $\pi 2^{3/2}$ is tight (see equation \eqref{leq13}
with $\alpha\rightarrow 0$ and $T\rightarrow 1$).
\EREM

\section{Piecewise polynomial functions}\label{z5}

Some kernels used in machine learning (e.\,g., the Epanechnikov kernel, see~\cite{DG01}, p.85)
are piecewise polynomial. Thus it is of interest to extend the result from the previous
section to higher-degree polynomials.

For $d>1$ we do not know how to sample from distribution ${\rm CI}_d(0,1)$ exactly. However we can still
approximately sample from this distribution, as follows. Let $r$ be an integer. Let $Z_1,\dots,Z_r$
be independent from Cauchy distribution $C(0,1/r)$. Consider the following distribution,
which we call $r$-approximation of ${\rm CI}_d(0,1)$:
\begin{equation}\label{apd}
(X_0,\dots,X_d)=\sum_{j=1}^r Z_j\cdot (1,(j/r),(j/r)^2,\dots,(j/r)^d).
\end{equation}
Now we show that if $r$ is large enough then the distribution given by~\eqref{apd} can be used instead of
distribution ${\rm CI}_d(0,1)$ for our purpose. As a consequence we will obtain the following.

\BTHM\label{trwe}
Let ${\cal F}$ consist of $m$ piecewise degree-$d$-polynomial densities, each consisting of $n$
pieces (given as in Remark~\ref{det}). We can obtain
$(\delta,\eps)$-relative-error approximation of $L_1$-distances between all
pairs in~${\cal F}$, using $O(m(m+n) d^3\eps^{-3}\ln(m/\delta))$ arithmetic operations.
\ETHM

\BREM
Note that for $d=1$ Theorem~\ref{trwe} gives worse (in $\eps$) running time that
Theorem~\ref{tele}. This slowdown is caused by the additional integration used to
simulate ${\rm CI}_d(0,1)$.
\EREM

The proof of Theorem~\ref{trwe} will be based on the following result which shows that
\eqref{apd} is in some sense close to ${\rm CI}_d(0,1)$.

\BLEM\label{lesa}
Let $p=a_0+a_1x+\dots+a_dx^d$ be a polynomial of degree $d$. Let $(X_0,\dots,X_d)$
be sampled from the distribution given by~\eqref{apd}, with $r\geq cd^2/\eps$ (where
$c$ is an absolute constant). Let
$W=a_0X_0+\dots+a_dX_d$. Then $W$ is from the Cauchy distribution $C(0,R)$, where
\begin{equation}\label{zwer}
(1-\eps)\int_{0}^1 |p(x)|\,{\rm d}x\leq R\leq (1+\eps)\int_{0}^1 |p(x)|\,{\rm d}x.
\end{equation}
\ELEM

We defer the proof of Lemma~\ref{lesa} to the end of this section. Note that
having \eqref{zwer} instead of \eqref{mrch} (which sampling from ${\rm CI}_d(0,1)$ would yield)
will introduce small relative error to the approximation of the $L_1$-distances.

\vskip 0.2cm
\BPRF[of Theorem~\ref{trwe}:]
The proof is analogous to the proof of Lemma~\ref{lpiece}. Let $r\geq cd^2/\eps$.
For $\ell\in\{1,\dots,s-1\}$ let $Z_\ell$ be independent from
$r$-approximation of ${\rm CI}_d(a_\ell,a_{\ell+1})$ distribution.
Let
$Y_{\ell}=Z_1+\dots+Z_{\ell-1}$, for $\ell=1,\dots,s$. Finally, for each $j\in [m]$, let
\begin{equation*}\label{stint2B}
X_j:=\sum_{\ell=1}^n \alpha_{j\ell}\cdot (Y_{c_{j\ell}} - Y_{b_{j\ell}})=\sum_{\ell=1}^n\alpha_{j\ell}
\cdot (Z_{b_{j\ell}}+\dots+Z_{c_{j\ell}-1}).
\end{equation*}
By Lemma~\ref{lesa}, for every $j,k\in [m]$ we have that $X_j-X_k$ is from Cauchy distribution $C(0,R)$ where
$(1-\eps)\|f_j-f_k\|_1\leq R\leq (1+\eps)\|f_j-f_k\|_1$.

If we have $t\geq (8/\eps)^2\ln (m^2/\delta)$ samples from each $X_1,\dots,X_m$ then using
Lemma~\ref{l:lhctail} and union bound with probability $\geq 1-\delta$ we recover
all $\|f_j-f_k\|_1$ with relative error $\approx 2\eps$.

Note that $s\leq 2mn$ and hence for the $Z_\ell$ we used $\leq 2mnt$ samples from
r-approximation of ${\rm CI}(0,1)$ distribution, costing us $O((d^3/\eps) mnt)$ arithmetic operation.
Computing the $Y_\ell$ takes $O(mnt)$ operations. Computing the $X_j$ takes $O(mnt)$ operations.
The final estimation of the distances takes $O(m^2 t)$ operations.
\EPRF

To prove Lemma~\ref{lesa} we will use the following Bernstein-type inequality from~\cite{Erd00}.

\BTHM\label{l:erdelyi}
(Theorem 3.1 of~\cite{Erd00})
There exists a constant $c>0$ such that for any degree $d$ polynomial $p$,
\[ \int_{0}^{1} | p'(x) | \mbox{d} x \leq c d^2 \int_0^1 | p(x) | \mbox{d} x. \]
\ETHM

We have the following corollary of Theorem~\ref{l:erdelyi}.

\BLEM\label{l:interpolate}
There exists a constant $c$ such that for any polynomial $p$ of degree $d$,
any $r \geq cd^2$, any $0 = x_0 < x_1 < x_2, \ldots < x_t = 1$ with $\max_{j} |x_j - x_{j-1}| \leq 1/r$,
and any $\theta_1 \in [x_0, x_1], \theta_2 \in [x_1, x_2], \ldots, \theta_{t} \in [x_{t-1},x_t]$, we have
\begin{equation}\label{e:interpolate}
(1 - cd^2/r) \int_0^1 | p(x) | \mbox{d} x \leq \sum_{j=1}^{t} (x_j - x_{j-1}) | p(\theta_j) | \leq
(1 + cd^2/r) \int_0^1 | p(x) | \mbox{d} x.
\end{equation}
\ELEM

\BPRF
We will use induction on the degree $d$ of the polynomial. For $d=0$ the sum and the integrals in \eqref{e:interpolate}
are equal.

Now assume $d\geq 1$. For each $j \in [t]$, we use the Taylor expansion of $p(x)$ about $\theta_j$ for
$x \in (x_{j-1}, x_{j}]$. This yields for each $x \in(x_{j-1}, x_j], ~
p(x) = p(\theta_j) + (x - \theta_j) p'(\theta'_{j,x})$, where $\theta'_{j,x} \in (x_{j-1}, x_j]$.
Let $\beta_j$ be the point $y\in (x_{j-1}, x_{j}]$ that maximizes $p'(y)$. We have
\begin{equation}\label{e:interpolate1}
\begin{split}
\left | \sum_{j=1}^{t} (x_j - x_{j-1}) | p(\theta_j) | - \int_0^1 | p(x) | \,\mbox{d} x
\right | \leq  \sum_{j=1}^{t} \int_{x_{j-1}}^{x_j} | p(x) - p(\theta_j) | \,\mbox{d} x\\
\leq \sum_{j=1}^{t} \int_{x_{j-1}}^{x_j} | (x - \theta_j) p'(\theta'_{j,x}) | \,\mbox{d} x
\leq \frac{1}{2r} \sum_{j=1}^{t} (x_j - x_{j-1}) |p'(\beta_{j})|.
\end{split}
\end{equation}
Since $p'$ is of degree $d - 1$, by induction hypothesis the right-hand side of~\eqref{e:interpolate1}
is bounded as follows
\begin{equation*}
\begin{split}
\frac{1}{2r} \sum_{j=1}^{t} (x_j - x_{j-1}) |p'(\beta_{j})|\leq
\frac{1}{2r} (1 + c(d-1)^2\eps) \int_{0}^{1} | p'(x) | \mbox{d} x \\
\leq (1/r) \int_{0}^{1} | p'(x) | \mbox{d} x \leq (cd^2/r) \int_{0}^{1} | p(x) | \mbox{d} x.
\end{split}
\end{equation*}
where in the last inequality we used Theorem~\ref{l:erdelyi}. Hence the lemma follows.
\EPRF

\BPRF[of Lemma~\ref{lesa}:]
We have
$$
W=(a_0,\dots,a_d)\cdot \sum_{j=1}^r Z_j (1,(j/r),(j/r)^2,\dots,(j/r)^d)=\sum_{j=1}^r Z_j\, p(j/r),
$$
where $Z_j$ are from Cauchy distribution $C(0,1/r)$. Thus $W$ is from Cauchy distribution $C(0,R)$, where
$$
R=\frac{1}{r}\sum_{j=1}^r |p(j/r)|.
$$
Using Lemma~\ref{l:interpolate} we obtain \eqref{zwer}.
\EPRF

\BREM
An alternate view of Lemma~\ref{l:interpolate} is that a piecewise degree-$d$-polynomial
density with $n$ pieces can be approximated by a piecewise uniform density with
$O(nd^2/\eps)$ pieces. The approximation distorts $L_1$-distances between any pair
of such densities by a factor at most $1\pm\eps$. To obtain a relative-approximation
of the $L_1$-distances in a family ${\cal F}$ one can now directly use the algorithm from Section~\ref{z2} without
going through the stochastic integrals approach (for $d=1$ the price for this
method is a $1/\eps$ factor slowdown).
\EREM

\BREM{\bf (on $L_2$-distances)}
For $L_2$-distances the dimension reduction uses normal distribution instead of Cauchy distribution.
For infinitesimal intervals the corresponding process is Brownian
motion, which is much better understood than Cauchy motion. Evaluation of a stochastic
integral of a deterministic function $\R\rightarrow\R^d$ w.r.t. Brownian motion is a
$d$-dimensional gaussian (whose covariance matrix is easy to obtain), for example
$$
\int_0^1(1,x,\dots,x^{d})\,{\rm d}{\cal L}_{\rm Brown}(x)
$$
is from $N(0,\Sigma)$ where $\Sigma$ is the $(d+1)\times (d+1)$ Hilbert matrix (that
is, the $ij$-th entry of $\Sigma$ is $1/(i+j-1)$).
\EREM

\BQUE
How efficiently can one sample from ${\rm CI}_d(0,1)$ distribution? A reasonable
guess seems to be that one can sample from a distribution within $L_1$-distance
$\delta$ from ${\rm CI}_d(0,1)$ using $d^2\ln(1/\delta)$ samples.
\EQUE

\vskip 0.2cm
\noindent
{\bf\Large Acknowledgement}

\vskip 0.2cm
\noindent
The authors would like to thank Carl Mueller for advice on stochastic integrals.

\bibliographystyle{alpha}
\bibliography{pairwisel1}

\section{Appendix}

\subsection{Stochastic integral of (constant, linear) function }

In this section we give an explicit formula for the density
function of the random variable
$$(X,Y)=\int_0^1\phi(z)\,{\rm d}{\cal L}(z),$$
where $\phi(z)=(1,z)$, and ${\rm d}{\cal L}(z)$ is the Cauchy
motion.

We will obtain the density function from the characteristic
function. The following result will be used in the inverse Fourier
transform. (We use $\Re$ to denote the real part of a complex
number.)

\BLEM\label{leee}
Let $\phi=(\phi_1,\dots,\phi_n):{\mathbb
R}\rightarrow {\mathbb R}^n$. Let
\begin{equation*}
Z=(X_1,\dots,X_n)=\int_{0}^1 \phi(x)\,\,{\rm d}{\cal L}(x),
\end{equation*}
where ${\cal L}$ is the Cauchy motion. The density function $f$ of $Z$ is given by
\begin{equation}\label{eeeeq}
\Re\(\frac{(n-1)!}{(2\pi)^n}
\int_{-\infty}^\infty\dots\int_{-\infty}^\infty
\frac{2}{(A+iB)^n}
{\rm d}b_1 \dots {\rm d}b_{n-1}\),
\end{equation}
where
\begin{equation}\label{zzz}
A=A(b_1,\dots,b_{n-1}):=\int_0^1 \left| b_1\phi_1(x)+\dots+b_{n-1}\phi_{n-1}(x) + \phi_n(x) \right|,
\end{equation}
and
\begin{equation}\label{zzzb}
B=B(b_1,\dots,b_{n-1},x_1,\dots,x_n):=b_1x_1 + \dots + b_{n-1} x_{n-1} + x_n.
\end{equation}
\ELEM

\BPRF The characteristic function of $Z$ is (see, e.\,g.,
proposition 3.2.2 of ~\cite{ST94}):
\begin{equation*}
\begin{split}
\hat{f}(a_1,\dots,a_n)=E[\exp(i(a_1X_1+\dots+a_nX_n))]=
\exp\(-\int_0^1 \left| a_1\phi_1(x)+\dots+a_n\phi_n(x) \right|\).
\end{split}
\end{equation*}
We will use the following integral, valid for any $A>0$ (see,
e.\,g.,~\cite{GR07}):
\begin{equation}\label{intog}
\int_0^\infty t^{n-1} \exp(-At) \cos(Bt)\,{\rm d}t = \frac{(n-1)!}{2}\(\frac{1}{(A-iB)^n}+\frac{1}{(A+iB)^n}\).
\end{equation}
We would like to compute the inverse Fourier transform of $\hat{f}$, which, since
$\hat{f}$ is symmetric about the origin, is given by
\begin{equation}\label{ew}
f(x_1,\dots,x_n)=
\frac{2}{(2\pi)^n}\int_{0}^\infty\int_{-\infty}^\infty\dots\int_{-\infty}^\infty
\hat{f}(a_1,\dots,a_n)\cos(a_1x_1+\dots+a_nx_n)
{\rm d}a_1 \dots {\rm d}a_{n-1} {\rm d}a_n.
\end{equation}
Substitution $a_n=t,a_{n-1}=b_{n-1}t,\dots,a_1=b_1t$ into \eqref{ew} yields
\begin{equation*}
\begin{split}
f(x_1,\dots,x_n)=\phantom{XXXXXXXXXXXXXXXXXXXXXXXXXXXXXXX}\\
\frac{2}{(2\pi)^n}
\int_{-\infty}^\infty\dots\int_{-\infty}^\infty
\Bigg(\int_{0}^\infty
t^{n-1}\exp\(-t \int_0^1 \left| b_1\phi_1(x)+\dots+b_{n-1}\phi_{n-1}(x) + \phi_n(x) \right|\)
\\
\cos\(t(b_1x_1 + \dots + b_{n-1} x_{n-1} + x_n)\)
{\rm d}t\Bigg)
{\rm d}b_1 \dots {\rm d}b_{n-1}.
\end{split}
\end{equation*}
Note that the inner integral has the same form as \eqref{intog} and hence
we have
\begin{equation}\label{ewr}
\begin{split}
f(x_1,\dots,x_n)=
\frac{(n-1)!}{(2\pi)^n}
\int_{-\infty}^\infty\dots\int_{-\infty}^\infty
\frac{1}{(A-iB)^n}+\frac{1}{(A+iB)^n}
{\rm d}b_1 \dots {\rm d}b_{n-1}\\
=\Re\(\frac{(n-1)!}{(2\pi)^n}
\int_{-\infty}^\infty\dots\int_{-\infty}^\infty
\frac{2}{(A+iB)^n}
{\rm d}b_1 \dots {\rm d}b_{n-1}\),
\end{split}
\end{equation}
where $A$ and $B$ are given by \eqref{zzz} and \eqref{zzzb}. The
last equality in \eqref{ewr} follows from the fact that the two
summands in the integral are conjugate complex numbers.
\EPRF

Now we apply Lemma~\ref{leee} for the case of two functions,
one constant and one linear.
\vskip 0.3cm

\BPRF[of Theorem~\ref{looo}:] Plugging $n=2$, $\phi_1(x)=1$, and
$\phi_2(x)=x$ into \eqref{zzz} and \eqref{zzzb} we obtain
\begin{equation}\label{eqB}
B(b_1,x_1,x_2)=b_1x_1+x_2
\end{equation}
and
\begin{equation}\label{eqA}
A(b_1)=\Bigg\{\begin{array}{rl}
b_1+1/2&\mbox{if}\ b_1\geq 0,\\
-b_1-1/2&\mbox{if}\ b_1\leq -1,\\
b_1^2+b_1+1/2&\mbox{otherwise}.\\
\end{array}
\end{equation}
Our goal now is to evaluate the integral \eqref{eeeeq}. We split the integral into
$3$ parts according to the behavior of $A(b_1)$.

We will use the following integral
\begin{equation}\label{nein1}
\int \frac{1}{(Sz + T)^2}\,{\rm d}z = - \frac{1}{S(T+Sx)}.
\end{equation}
For $B=b_1x_1+x_2$ and $A=b_1+1/2$ we have $A+iB = b_1(1+ix_1)+(1/2+ix_2)$. Using
\eqref{nein1} for $A$ and $B$ given by \eqref{eqB} and \eqref{eqA}) we obtain
\begin{equation}\label{e1xx}
\int_{0}^\infty
\frac{1}{(A+iB)^2}{\rm d}b_{1}=\frac{2}{(i x_1 + 1)(2 i x_2 + 1)},
\end{equation}
and
\begin{equation}\label{e3xx}
\int_{-\infty}^{-1}
\frac{1}{(A-iB)^2}
{\rm d}b_{1}=
\frac{2}{(i x_1 - 1)(2i(x_1 - x_2) - 1)}.
\end{equation}
We have (see, e.\,g.,~\cite{GR07}))
\begin{equation}\label{nein}
\int \frac{1}{(z^2 + Sz + T)^2}\,{\rm d}z =  \frac{S + 2 z}{(4 T
-S^2) (T + S z + z^2)} + \frac{
 4 {\rm atan} \((S + 2 z)/\sqrt{ 4 T -S^2}\)}{(4 T -S^2)^{3/2}}.
\end{equation}
For $A=b_1^2+b_1+1/2$ and $B=b_1x_1+x_2$ we have $A+iB = b_1^2 + b_1 ( 1 + ix_1 ) + (1/2+x_2)$.
Using \eqref{nein} we obtain
\begin{equation}\label{moui}
\int_{-1}^0
\frac{1}{(A+iB)^2}{\rm d}b_{1}
=
\frac{2 (i x_1+1)}{(2 i x_2+1)Q}+
\frac{2 (i x_1-1)}{(2 i (x_1-x_2)-1)Q}+
4\frac{{\rm atan}\(\frac{i x_1+1}{\sqrt{Q}}\)-{\rm atan}\(\frac{i x_1-1}{\sqrt{Q}}\)}
{Q^{3/2}},
\end{equation}
where $Q$ is given by \eqref{qeq}.

Summing \eqref{e1xx}, \eqref{e3xx}, and \eqref{moui} we obtain
\begin{equation}\label{ekrt}
\int_{-\infty}^\infty
\frac{1}{(A+iB)^2}{\rm d}b_{1}=\frac{8}{Q(1+x_1^2)}
+4\frac{{\rm atan}\(\frac{i x_1+1}{\sqrt{Q}}\)-{\rm atan}\(\frac{i x_1-1}{\sqrt{Q}}\)}
{Q^{3/2}}.
\end{equation}
We have
$$
\left|\frac{ix_1\pm 1}{\sqrt{Q}}\right|^4=\frac{(1+x_1^2)^2}{(1 + x_1^2)^2 + (2x_1-4x_2)^2}\leq 1.
$$
with equality only if $x_1=2x_2$. Hence if $x_1\neq 2x_2$ then using \eqref{eeer2} we have
\begin{equation*}
{\rm atan}\(\frac{i x_1+1}{\sqrt{Q}}\)-{\rm atan}\(\frac{i x_1-1}{\sqrt{Q}}\)={\rm atan}(iQ/(x_1-2x_2)),
\end{equation*}
and by applying
\begin{equation*}\label{uhy}
\Re\(\frac{8}{Q(1+x_1^2)}\)=\frac{8}{1 + 6 x_1^2 + x_1^4 - 16 x_1 x_2 + 16 x_2^2}
\end{equation*}
in \eqref{ekrt} we obtain \eqref{eden}.

If $x_1=2x_2$ then $Q=1+x_1^2$ and using
\begin{equation*}\label{pppq}
{\rm atan}\(\frac{i x_1+1}{\sqrt{Q}}\)-{\rm atan}\(\frac{i x_1-1}{\sqrt{Q}}\)=\pi/2
\end{equation*}
in \eqref{ekrt} we obtain \eqref{eden2}.
\EPRF

\subsection{Bounding the ${\rm CI}_1(0,1)$-distribution}\label{sbound}

Now we prove that the multivariate student distribution gives
an efficient envelope for the ${\rm CI}_1(0,1)$-distribution.

\vskip 0.2cm
\BPRF[of Lemma~\ref{kkkk}:]
To simplify the formulas we use the following substitutions: $x_1=u$ and $x_2=w+u/2$.
The density $g$ becomes
\begin{equation*}\label{ZZZ2}
g'(u,v):=\frac{1}{\pi}\(1+u^2+4w^2\)^{-3/2}.
\end{equation*}
For $w=0$ (which corresponds to $x_1=2x_2$) the density $f$ becomes
\begin{equation}\label{eden2B}
\frac{4/\pi^2}{(1 + u^2)^2}
+\frac{1}{\pi (1+u^2)^{3/2}},
\end{equation}
and hence Lemma~\ref{kkkk} is  true, as
$$
\eqref{eden2B}\leq (4/\pi+1)\(\frac{1}{\pi}\(1+u^2\)^{-3/2}\).
$$

For $w\neq 0$, density \eqref{eden} becomes
\begin{equation*}\label{zzzq}
f'(u,v):=\frac{1}{\pi^2}\(
\frac{4}{(1+u^2)^2+(4w)^2}+\frac{{\rm atan}(iM/(2w))}{M^{3}}-\frac{{\rm atan}(iM'/(2w))}{{M'}^{3}}\),
\end{equation*}
where $M=(1+u^2-4iw)^{1/2}$ and $M'=(1+u^2+4iw)^{1/2}$. We are going to show
\begin{equation}\label{leq2}
\pi^2 f'(u,v)
\leq C\pi g'(u,v).
\end{equation}

Note that both sides of \eqref{leq2} are unchanged when we flip the sign of $u$ or the sign of $w$.
Hence we can, without loss of generality, assume $u\geq 0$ and $w>0$.

There are unique $a>0$ and $b>0$ such that $w=ab/2$ and $u=\sqrt{a^2-b^2-1}$ (to see this
notice that substituting $b=2w/a$ into the second equation yields $u^2+1 = a^2 - 4w^2/a^2$,
where the right-hand side is a strictly increasing function going from $-\infty$ to $\infty$).
Note that $M=a-ib$ and $M'=a+ib$. Also note that
\begin{equation}\label{econs}
a^2\geq b^2 + 1.
\end{equation}
After the substitution equation \eqref{leq2} simplifies as follows
\begin{equation}\label{leq4}
\begin{split}
\frac{4}{(a^2+b^2)^2}+\frac{1}{(a^2+b^2)^3}\Bigg(
(a+ib)^3\,{\rm atan}\(\frac{1}{a}+\frac{i}{b}\)\phantom{XXXXXXXXXXXXXXX}\\
+(a-ib)^3\,{\rm atan}\(\frac{1}{a}-\frac{i}{b}\)
\Bigg)
\leq \frac{C}{(a^2-b^2+a^2b^2)^{3/2}}.
\end{split}
\end{equation}
Now we expand $(a+ib)^3$ and $(a-ib)^3$ and simplify \eqref{leq4} into
\begin{equation}\label{leq5}
\begin{split}
\frac{4}{(a^2+b^2)^2}+\frac{1}{(a^2+b^2)^3}\Bigg(
(a^3-3ab^2)\({\rm atan}\(\frac{1}{a}+\frac{i}{b}\)+{\rm atan}\(\frac{1}{a}-\frac{i}{b}\)\)\phantom{XXXXXX}\\
-i(b^3-3a^2b)\({\rm atan}\(\frac{1}{a}+\frac{i}{b}\)-{\rm atan}\(\frac{1}{a}-\frac{i}{b}\)\)
\Bigg)
\leq \frac{C}{(a^2-b^2+a^2b^2)^{3/2}}.
\end{split}
\end{equation}
Now we substitute $a=1/A$ and $b=1/B$ into \eqref{leq5} and obtain
\begin{equation}\label{leq6}
\begin{split}
\frac{4A^4B^4}{(A^2+B^2)^2}+\frac{A^3B^3}{(A^2+B^2)^3}\Bigg(
(B^3-3A^2B)\big({\rm atan}\(A+iB\)+{\rm atan}\(A-iB\)\big)\phantom{XXXXXXX}\\
-i(A^3-3AB^2)\big({\rm atan}\(A+iB\)-{\rm atan}\(A-iB\)\big)
\Bigg)
\leq \frac{C\cdot A^3B^3}{(B^2-A^2+1)^{3/2}}.
\end{split}
\end{equation}
Note that $A>0$ and $B>0$ and the constraint \eqref{econs} becomes
\begin{equation}\label{econs2}
B^2\geq A^2(1+B^2).
\end{equation}
Multiplying both sides of \eqref{leq6} by $(A^2+B^2)^3/(AB)^3$ we obtain
\begin{equation}\label{leq7}
\begin{split}
4AB(A^2+B^2)+
(B^3-3A^2B)\big({\rm atan}\(A+iB\)+{\rm atan}\(A-iB\)\big)\phantom{XXXXXXXXX}\\
-i(A^3-3AB^2)\big({\rm atan}\(A+iB\)-{\rm atan}\(A-iB\)\big)
\leq \frac{C\cdot (A^2+B^2)^6}{(B^2-A^2+1)^{3/2}}.
\end{split}
\end{equation}
Finally, we substitute $A=T\sin\alpha$ and $B=T\cos\alpha$ with $T\geq 0$.
Note that the constraint \eqref{econs2} becomes
\begin{equation}\label{econs3}
(T\sin\alpha)^2\leq \frac{\cos(2\alpha)}{(\cos\alpha)^2},
\end{equation}
and hence $\alpha$ is restricted to $[0,\pi/4)$.

Equation \eqref{leq7} then becomes
\begin{equation}\label{leq8}
\begin{split}
2T^4\sin(2\alpha)+
T^3\cos(3\alpha)\({\rm atan}\(A+iB\)+{\rm atan}\(A-iB\)\)\phantom{XXXXXXXXXX}\\
+iT^3\sin(3\alpha)\({\rm atan}\(A+iB\)-{\rm atan}\(A-iB\)\)
\leq \frac{C\cdot T^6}{(T^2\cos(2\alpha)+1)^{3/2}}.
\end{split}
\end{equation}
We prove~\eqref{leq8} by considering three cases.

\vskip 0.2cm
\underline{{\bf CASE: $T<1$}}. We can use \eqref{eeer2} to simplify~\eqref{leq8} as follows
\begin{equation}\label{leq9}
\begin{split}
2T\sin(2\alpha)+
\cos(3\alpha)\,{\rm atan}\(\frac{2T\sin(\alpha)}{1-T^2}\)
-\sin(3\alpha)\,{\rm atanh}\(\frac{2T\cos(\alpha)}{1+T^2}\)
\leq \frac{C\cdot T^3}{(T^2\cos(2\alpha)+1)^{3/2}}.
\end{split}
\end{equation}
For $z\geq 0$ we have ${\rm atanh}(z)\geq z\geq{\rm atan}(z)$ and hence to prove~\eqref{leq9}
it is enough to show
\begin{equation}\label{leq10}
\begin{split}
2T\sin(2\alpha)(1-T^4)+
(1+T^2)\cos(3\alpha)\,\(2T\sin(\alpha)\)\phantom{XXXXXXXXX}\\
-(1-T^2)\sin(3\alpha)\,\(2T\cos(\alpha)\)
\leq \frac{C\cdot T^3(1-T^4)}{(T^2\cos(2\alpha)+1)^{3/2}},
\end{split}
\end{equation}
which is implied by the following inequality which holds for $T\leq 8/9$:
\begin{equation}\label{leq11}
-2T^2\sin(2\alpha)+2\sin(4\alpha)
\leq 2 \leq \frac{C\cdot 2465/6561}{2^{3/2}} \leq \frac{C\cdot (1-T^4)}{(T^2\cos(2\alpha)+1)^{3/2}}.
\end{equation}
For $1>T\geq 8/9$ we directly prove~\eqref{leq10}
\begin{equation*}\label{leq12}
\begin{split}
2T\sin(2\alpha)+
\cos(3\alpha)\,{\rm atan}\(\frac{2T\sin(\alpha)}{1-T^2}\)
-\sin(3\alpha)\,{\rm atanh}\(\frac{2T\cos(\alpha)}{1+T^2}\)\\
\leq 2+\pi/2 \leq \frac{C\cdot 512/729}{2^{3/2}} \leq \frac{C\cdot T^3}{(T^2\cos(2\alpha)+1)^{3/2}}.
\end{split}
\end{equation*}

\underline{{\bf CASE: $T>1$}.} We can use \eqref{eeer3} and \eqref{lh1} to simplify~\eqref{leq8} as follows
\begin{equation}\label{leq13}
\begin{split}
2T\sin(2\alpha)+
\cos(3\alpha)\(\pi+{\rm atan}\(\frac{2T\sin(\alpha)}{1-T^2}\)\)
-\phantom{XXXXXXXXXXX}\\
\sin(3\alpha)\,{\rm atanh}\(\frac{2T\cos(\alpha)}{1+T^2}\)
\leq \frac{C\cdot T^3}{(T^2\cos(2\alpha)+1)^{3/2}}.
\end{split}
\end{equation}
From \eqref{econs3} we have $T\sin(\alpha)\leq 1$ and hence $2T\sin(2\alpha)\leq 4$. Therefore
\eqref{leq13} can be proved as follows.
\begin{equation*}\label{leq14}
\begin{split}
2T\sin(2\alpha)+
\cos(3\alpha)\(\pi+{\rm atan}\(\frac{2T\sin(\alpha)}{1-T^2}\)\)
-\sin(3\alpha)\,{\rm atanh}\(\frac{2T\cos(\alpha)}{1+T^2}\)\phantom{XXXXXX}\\
\leq 4+3\pi/2\leq \frac{C}{2^{3/2}}\leq\frac{C\cdot T^3}{(T^2\cos(2\alpha)+1)^{3/2}}.
\end{split}
\end{equation*}

\underline{{\bf CASE: $T=1$}.}
Equation \eqref{leq8} simplifies as follows
\begin{equation}\label{leq15}
\begin{split}
2\sin(2\alpha)+(\pi/2)\cos(3\alpha)-\sin(3\alpha)\,{\rm atanh}\(\cos(\alpha)\)
\leq \frac{C}{(\cos(2\alpha)+1)^{3/2}}.
\end{split}
\end{equation}
The left-hand side is bounded from above by $2+\pi/2$ which is less than $C/2^{3/2}$
which lower-bounds the right-hand side of \eqref{leq15}.
\EPRF

\subsection{Basic properties of trigonometric functions}

In this section we list the basic properties of trigonometric
functions that we used. For complex parameters these are
multi-valued functions for which we choose the branch in the
standard way. The {\em logarithm} of a complex number
$z=(\cos\alpha + i\sin\alpha){\mathrm e}^t$, where $\alpha\in
(-\pi,\pi]$, and $t\in\R$ is $i\alpha+t$. The {\em inverse
tangent} of a complex number $z\in{\mathbb C}\setminus\{\pm i\}$
is the solution of $\tan(x)=z$ with $\Re(x)\in(-\pi/2,\pi/2)$. In
terms of the logarithm we have
\begin{equation*}\label{atan}
{\rm atan}(z):=\frac{1}{2}i\(\ln(1-iz)-\ln(1+iz)\).
\end{equation*}
The inverse hyperbolic tangent function is defined analogously, for
$z\in{\mathbb C}\setminus\{\pm 1\}$ we have
\begin{equation*}\label{atanh}
{\rm atanh}(z):=\frac{1}{2}\(\ln(1+z)-\ln(1-z)\)=-i\,{\rm atan}(iz).
\end{equation*}
For non-negative real numbers $z$ we have the following inequality
\begin{equation*}
{\rm atanh}(z)\geq z\geq{\rm atan}(z).
\end{equation*}
The ${\rm atan}$ function (even as a multi-valued function) satisfies
\begin{equation*}\label{eeer}
{\rm tan}({\rm atan}(x) + {\rm atan}(y))=\frac{x+y}{1-xy},
\end{equation*}
for any values of $x,y\in{\mathbb C}\setminus\{\pm i\}$, with $xy\neq 1$.

For $a^2+b^2<1$ the real part of ${\rm atan}(a+bi)$ is from $(-\pi/4,\pi/4)$. Hence
\begin{equation}\label{eeer2}
|x|<1\ \wedge |y|<1\ \implies\ {\rm atan}(x) + {\rm atan}(y)={\rm atan}\(\frac{x+y}{1-xy}\).
\end{equation}
For $a\geq 0$ and $a^2+b^2\geq 1$ the real part of ${\rm atan}(a+bi)$ is from $[\pi/4,\pi/2)$.
\begin{equation}\label{eeer3}
a\geq 0 \wedge a^2+b^2>1 \implies {\rm atan}(a+bi)+{\rm atan}(a-bi)=\pi+{\rm atan}(2a/(1-a^2-b^2)).
\end{equation}
For $a\geq 0$ the real part of ${\rm atan}(a+bi)$ is from $[0,\pi/2)$. Hence
for any $a,b$ with $a+ib\neq\pm i$ we have
\begin{equation}\label{lh1}
{\rm atan}(a+bi)-{\rm atan}(a-bi)={\rm atan}\(\frac{2ib}{1+a^2+b^2}\).
\end{equation}

\end{document}